\documentclass[aps,prl,reprint,twocolumn,citeautoscript,superscriptaddress,
showkeys,showpacs, letter]{revtex4-1}
\usepackage{xr}
\usepackage[fleqn,tbtags]{amsmath}
\usepackage{amsfonts, amssymb,amsxtra,textcomp}
\usepackage[]{graphicx}
\usepackage{grffile}
\usepackage{mathtools}
\usepackage{color}                                      % define colors
\definecolor{d_red}{cmyk}{0.00, 0.81, 1.00, 0.27}
\definecolor{d_orange}{cmyk}{0.00, 0.33, 1.00, 0.00}
\definecolor{d_blue}{cmyk}{0.78, 0.47, 0.00, 0.20}
\definecolor{d_lgreen}{cmyk}{0.07, 0.00, 0.79, 0.29}
\definecolor{d_green}{cmyk}{0.66, 0.00, 0.71, 0.56}
\definecolor{d_blue}{cmyk}{0.78, 0.47, 0.00, 0.20}
\definecolor{d_dblue}{cmyk}{0.91, 0.79, 0.00, 0.22}
\definecolor{d_pink}{cmyk}{0.0, 0.79, 0.37, 0.29}
\definecolor{d_purple}{cmyk}{0.16, 0.54, 0.00, 0.70}
\definecolor{d_paleblue}{cmyk}{0.669, 0.338, 0.00, 0.373}
\definecolor{d_dpaleblue}{cmyk}{0.441, 0.290, 0.00, 0.580}
\definecolor{d_brown}{cmyk}{0.0, 0.490, 0.930, 0.350}
\definecolor{d_turquoise}{cmyk}{0.630, 0.04, 0.0, 0.440}

\newcommand{\av}[1]{\langle #1 \rangle}
\newcommand{\bfss}{{\boldsymbol{S}}}

\newcommand{\bfn}{{\boldsymbol{n}}}

\newcommand{\bfb}{{\boldsymbol{b}}}

\newcommand{\bfh}{{\boldsymbol{h}}}
\newcommand{\bft}{{\boldsymbol{t}}}

\def\bmx{\begin{pmatrix}}
  \def\emx{\end{pmatrix}}

\usepackage{hyperref}
\usepackage[figure,table]{hypcap}               % to correct a problem with hyperref
\hypersetup{
  pdftitle = {},
  pdfsubject = {},
  pdfauthor = {},
  pdfkeywords = {},
  pdfcreator = {},
  pdfproducer = {LaTeX with hyperref},
  colorlinks = true,
  linkcolor = blue, 
  anchorcolor = blue,
  citecolor = blue, 
  filecolor = red, 
  menucolor = red, 
  pagecolor = red, 
  urlcolor  = blue, 
  breaklinks = true,
  pdfstartview = FitV,
  pdfhighlight = /I,
  pdfpagelayout = OneColumn,
  hypertexnames=true 
}

\begin{document}
\title{Emergent Critical Phase and Ricci Flow in a 2D Frustrated Heisenberg Model}
\author{Peter P. Orth}
% \email{peter.orth@kit.edu}
\affiliation{Institute for Theory of Condensed Matter, Karlsruhe Institute of Technology (KIT), 76131 Karlsruhe, Germany}
\author{Premala Chandra}
\affiliation{Center for Materials Theory,  Rutgers University, Piscataway, New Jersey 08854, USA}
% Department of Physics and Astronomy,
% Center for Materials Theory,%Rutgers University, Piscataway, NJ 08855, U.S.A.
\author{Piers Coleman}
\affiliation{Center for Materials Theory,  Rutgers University, Piscataway, New Jersey 08854, USA}
\affiliation{Hubbard Theory Consortium and Department of Physics, Royal Holloway, University of London, Egham, Surrey TW20 0EX, UK}
\author{J\"org Schmalian}
\affiliation{Institute for Theory of Condensed Matter, Karlsruhe Institute of Technology (KIT), 76131 Karlsruhe, Germany}
\affiliation{DFG Center for Functional Nanostructures, Karlsruhe Institute of Technology (KIT), 76128 Karlsruhe, Germany}
\date{\today}

\begin{abstract}
  We introduce a two-dimensional frustrated Heisenberg antiferromagnet
  on interpenetrating honeycomb and triangular lattices.  Classically
  the two sublattices decouple, and ``order from disorder'' drives them
  into a coplanar state. Applying Friedan's geometric approach to
  nonlinear sigma models, we obtain the scaling of the
  spin-stiffnesses governed by the Ricci flow of a 4D metric
  tensor. At low temperatures, the relative phase between the spins on
  the two sublattices is described by a six-state clock model with an
  emergent critical phase. 
\end{abstract}

% \keywords{keywords}

\maketitle
% According to the Hohenberg-Mermin-Wagner (HMW) theorem, 
% continuous symmetries cannot be broken at finite temperatures
% in two dimensional 
% systems with
% short-range interactions
% due to long wavelength thermal
% fluctuations. In 2D Heisenberg systems these fluctuations
% invariably lead to a finite spin correlation length. 

A remarkable discovery of recent years is that frustrated two dimensional Heisenberg models can evade the Hohenberg-Mermin-Wagner theorem~\cite{PhysRev.158.383,mermin_absence_1966} via the development of long-range discrete order driven by short-range thermal spin fluctuations: such discrete long-range order develops despite the persistence of a finite spin correlation length, leading to a finite temperature Ising ($\mathbb{Z}_{2}$) or $\mathbb{Z}_3$ Potts phase transition~\cite{PhysRevLett.64.88,PhysRevLett.92.157202,PhysRevLett.91.177202, PhysRevB.81.214419, annurev-conmatphys-070909-104138, PhysRevB.78.094423, arXiv:1207.4752v1}. 
This phenomenon is well-established in the $J_{1}$-$J_{2}$ Heisenberg model on the square lattice and has recently been realized in iron-based superconductors~\cite{PhysRevLett.105.157003}. An interesting question motivated by this discovery is whether it can be generalized to higher $\mathbb{Z}_p$ ($p\geq 5$) order. If one can show, in addition, that these emergent discrete degrees of freedom are described by a $p$-state clock model~\cite{PhysRevB.16.1217, Ortiz2012780}, the unique situation arises that a Heisenberg spin system exhibits two Berezinskii-Kosterlitz-Thouless (BKT) transitions which bracket a critical phase. In a system of discrete Ising spins, such a scenario was reported to occur on the triangular lattice~\cite{PhysRevB.29.5250, PhysRevB.68.104409}. 
\begin{figure}[b]
  \centering
  \includegraphics[width=.8\linewidth]{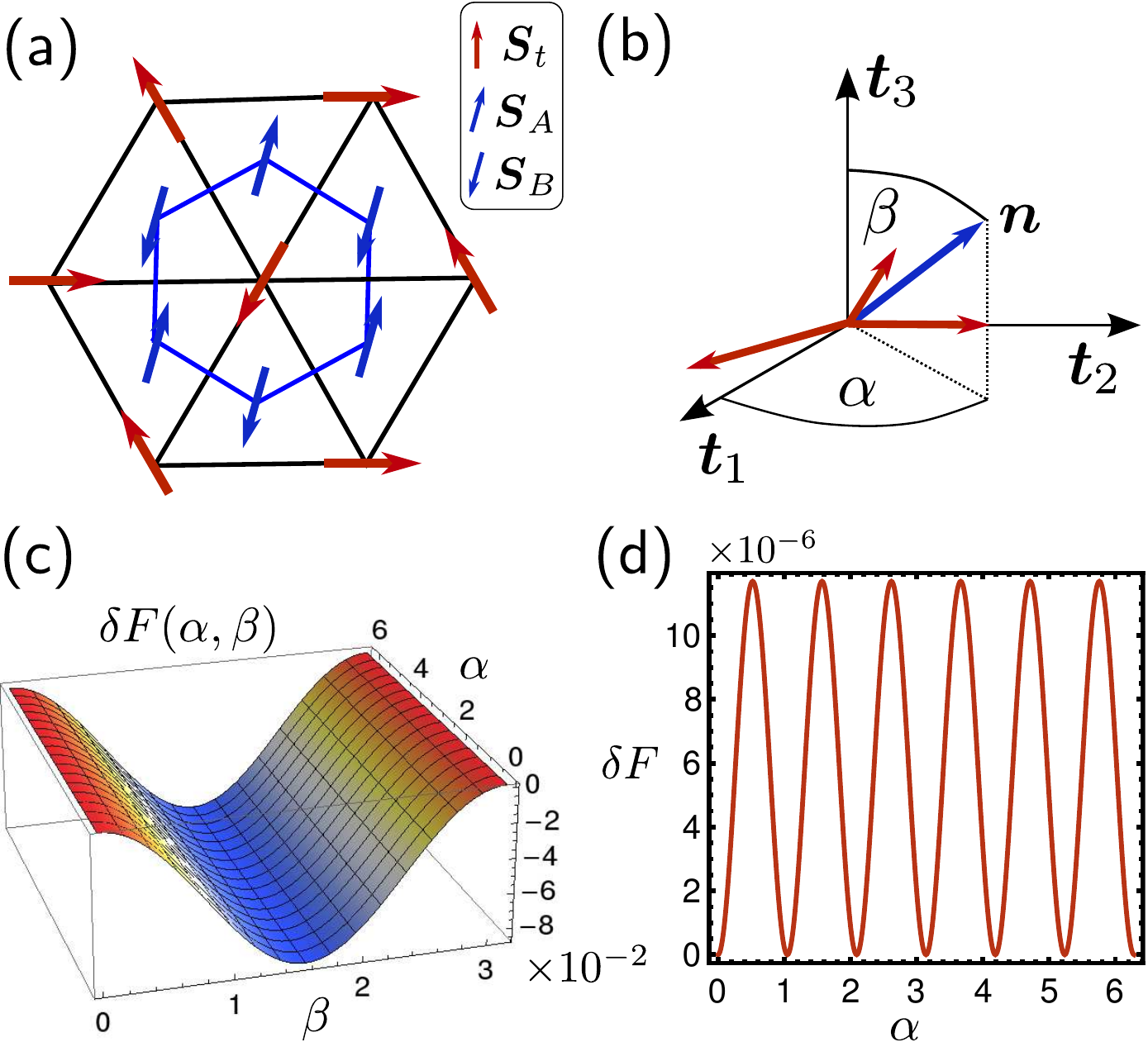}
  \caption{(color online) (a) Heisenberg model on the ``windmill'' lattice. (b) Definition of angles $\alpha$ and $\beta$ describing the relative orientation of magnetic order on triangular and honeycomb lattice. (c--d) Angle dependent free energy correction $\delta F$ from thermal and quantum spin fluctuations for parameters $J_{hh} = J_{tt} = 1$, $J_{th} = 0.4$, $T = 1$. Panel (d) is for fixed $\beta = \pi/2$. }
  \label{fig:1}
\end{figure}

In this Letter, we introduce such a Heisenberg model defined on interpenetrating honeycomb and triangular lattices (Fig.~\ref{fig:1}(a)) with nearest-neighbor antiferromagnetic coupling. This model may be realized with cold spinful atoms in optical lattices, where it arises naturally in the limit of large on-site interactions~\cite{PhysRevLett.91.090402, PhysRevLett.107.165301,Trotzky-SuperexchangeColdAtoms-Science-2008,PhysRevLett.107.210405}. %PhysRevA.81.061603,  bakr_quantum_2009, EsslingerDirac-Nature2012
Another promising experimental route is to employ STM techniques for nano-fabrication and spin-resolved read-out of stacked triangular and honeycomb monolayers of magnetic atoms like Cr or Co~\cite{manoharan-TwoKondoMirage-Nature-2000,Manoharan-ArtificialMolecularGraphene-Nature-2012,PhysRevLett.101.267205,PhysRevB.82.012402}. 
For classical spins the two sublattices are decoupled giving rise to an SO(3)$\times$O(3)$/$O(2) order parameter. ``Order from disorder''~\cite{villain-JPhysFrance-1977,PhysRevLett.62.2056} drives the two sublattices into a coplanar spin configuration~\cite{PhysRevLett.68.855} with an SO(3)$\times$U(1) order parameter and a six-fold
in-plane potential. In the coplanar state we explicitly show that the $U(1)$ degrees of freedom decouple to form an emergent $\mathbb{Z}_{6}$ clock model with an intermediate power-law phase. This non-trivial decoupling of the $U(1)$ phase is essential for the critical phase to occur.

% An aspect of our work  of possible
% that promises to be of 
% relevance beyond the immediate field of frustrated magnetism is 
% An innovation of the current work is
A novel aspect of our work is that we apply Friedan's coordinate-independent approach to nonlinear sigma models~\cite{PhysRevLett.45.1057} to the scaling of the spin-stiffness. In this approach the configurations of the 2D spin system correspond to a worldsheet of a string evolving in four-dimensions, where the metric is determined by the components of the antiferromagnetic stiffness and its renormalization corresponds to a Ricci flow of the metric tensor. We also note the decoupling of our $U(1)$ phase can be viewed as a toy model for the compactification of a four-dimensional string-theory. 
% into a one-dimensional manifold.

Specifically we study the antiferromagnetic Heisenberg model on a decorated 2D triangular lattice 
(cf. Fig.~\ref{fig:1}(a)); the associated Bravais lattice has three basis sites per unit cell at positions $\bfb_t = a_0 (0, 2/\sqrt{3})$,
$\bfb_A = (0,0)$ and $\bfb_B = a_0 (0, 1/\sqrt{3})$ where indices $A,B$ label the two honeycomb sites. We set the lattice constant $a_0=1$. The Hamiltonian is $H = H_{tt} + H_{AB} + H_{tA} + H_{tB}$ with
\begin{align}
  \label{eq:1}
  H_{ab} &= J_{ab} \sum_{j=1}^{N_L} \sum_{\{\delta_{ab}\}} \bfss_{a}(r_j) \cdot \bfss_b(r_j + \delta_{ab})\,,
\end{align}
where $\bfss_a(r_j)$ denote 
% quantum 
spin operators at Bravais
lattice site $j$ and basis site $a \in \{t, A,B \}$. The vectors
$\{\delta_{ab}\}$ point between nearest-neighbors of
sublattices $a,b$. %While it is not essential for the
%qualitative behavior of this model, 
We assume in the following that the spin exchange couplings within the same sublattice are larger than
the inter-sublattice coupling $J_{th} < J_{tt}, J_{hh}$, where
$J_{th} \equiv J_{tA} = J_{tB}$ and $J_{hh} \equiv J_{AB}$. For
decoupled lattices $J_{th} = 0$, the classical ground state on the
bipartite honeycomb lattice is the usual N\'eel state, while spins on
the triangular lattice arrange in a $120$\textdegree
configuration~\cite{sachdev_qpt_book}. Although the exchange fields
between the two sublattices exactly cancel for this configuration even
for $J_{th} > 0$, quantum and thermal fluctuations depend on the
relative orientation of the magnetization on the two sublattices. 
The uniaxial magnetic order on the honeycomb lattice is described by a
normal vector $\bfn(x)$, which points along the magnetization on
sublattice $A$. The biaxial order on the triangular lattice is
characterized by a triad
of orthonormal vectors $\{\bft_j(x)\}$ with $j=1,2,3$. Equivalently, it may be expressed by an orthogonal matrix $t = (\bft_1, \bft_2, \bft_3) \in SO(3)$. We take the vectors $\bft_{1,2}$ to span the plane of the magnetization on the triangular lattice.  The relative order between the two sublattices is thus determined by two angles $\alpha$ and $\beta$, that  are defined in Fig.~\ref{fig:1}(b). 
\begin{figure}[t!]
  \centering
  \includegraphics[width=.92\linewidth]{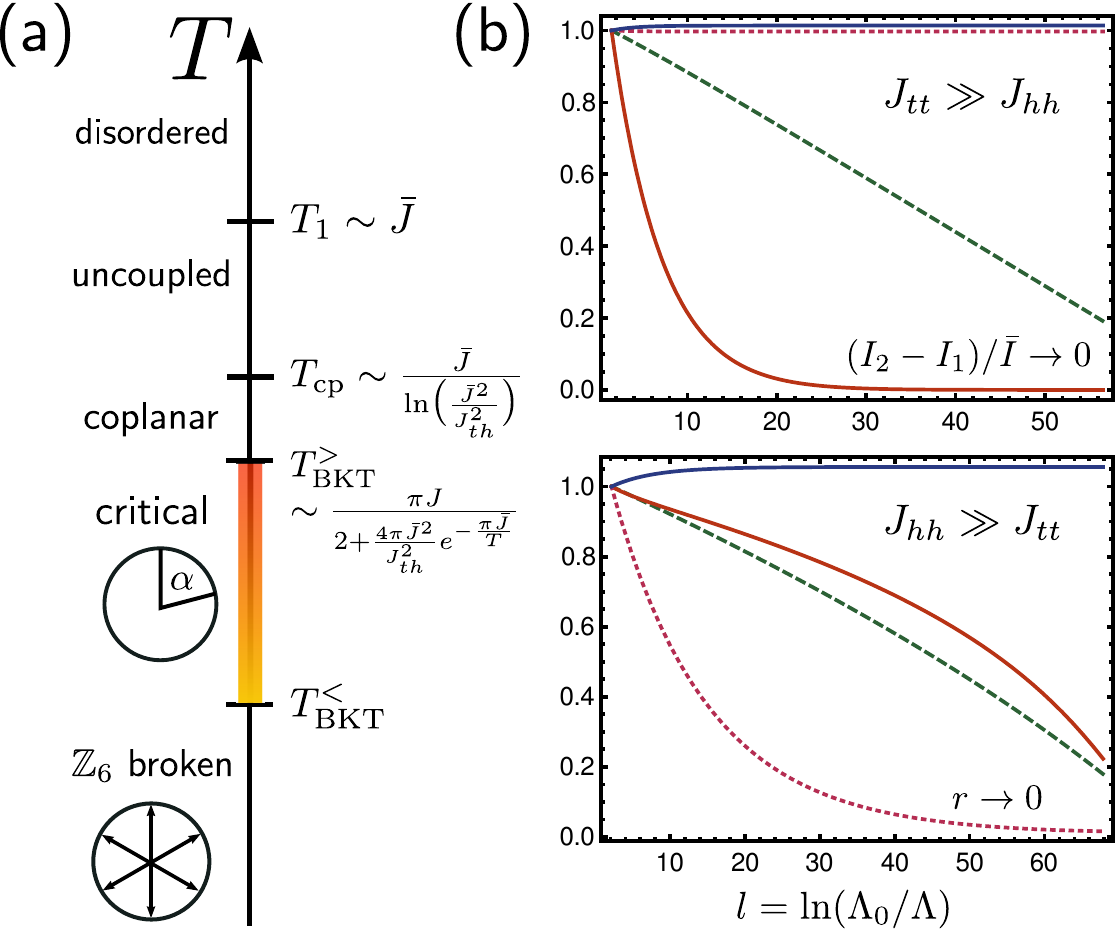}
  \caption{(a) Schematic phase diagram. (b) Coplanar RG flow of the variables $I'_\alpha$ (blue, increasing), $\bar{I} = (I_1 I_2 I_3)^{1/3}$ (greed dashed), $(I_2 - I_1)/\bar{I}$ (red), and $r$ (pink dotted). Curves are normalized to initial values at $l_\gamma$. Upper panel is for $J_{tt} \gg J_{hh}$ with $J_{hh} = 1$, $J_{tt} = 5$, $J_{th} = 0.4$, $T = 0.6$, and initial values $\bar{I}=5.3$, $(I_2 - I_1)/\bar{I}=0.27$, $r = 0.82$, $I'_\alpha = 1.2$. Decoupling is due to $(I_1 - I_2)/\bar{I} \rightarrow 0$. Lower panel is for $J_{hh} \gg J_{tt}$ with $J_{hh} = 5$, $J_{tt} = 1$, $J_{th} = 0.4$, $T = 0.5$, and initial values  $\bar{I}=4.5$, $(I_2 - I_1)/\bar{I}=2.1$, $r = 0.11$, $I'_\alpha = 1.1$. Decoupling is due to $r \rightarrow 0$.}
  \label{fig:3}
\end{figure}

% \emph{Long-wavelength action.--} 
Symmetry considerations dictate the form of the long-wavelength action which takes the form of a nonlinear sigma model (NLSM)
\begin{equation}
  \label{eq:2}
  S = \int d^2x \biggl( \frac{K}{2} (\partial_\mu \bfn)^2 + \sum_{j=1}^3 \frac{K_j}{2} (\partial_\mu \bft_j)^2 \biggr) +S_c \,.
\end{equation}
% \\ & + \frac{\gamma}{a^2} (\bfn \cdot \bft_3)^2 + \frac{\lambda}{a^2} \Bigl[ (\bfn \cdot \bft_2)^3 - 3 (\bfn \cdot \bft_2) (\bfn \cdot \bft_1)^2 \Bigr]^2 \biggr\}  \nonumber 
% where $a$ denotes the short-distance cutoff, that is initially equal to the lattice spacing $a_0$. 
The action contains the usual gradient terms of the $O(3)/O(2)$ and
the $SO(3)$ NLSM for the order parameter on the honeycomb and
triangular lattice. The bare spin stiffnesses $K, K_j$ can be derived
in a $1/S$-expansion and read $K = 2 J_{hh} S^2/T$, $K_1 = K_2 =
\sqrt{3} J_{tt} S^2/4 T$ and $K_3 = 0$~\cite{PhysRevLett.50.1153,
  PhysRevB.39.6797,PhysRevB.39.2344}. In addition, the action in
Eq.~\eqref{eq:2} contains two potential terms, generated by short-wavelength spin fluctuations (``order from disorder'')~\cite{villain-JPhysFrance-1977,PhysRevLett.62.2056}
% that
% depend on the
% relative orientation of the magnetic order on the two sublattices
\begin{equation}
  \label{eq:35}
  S_c=\frac12 \int d^{2}x\left( \gamma \cos ^{2}\beta +\lambda \sin ^{6}\beta \sin
    ^{2}\left( 3\alpha \right) \right) \,.
\end{equation}
% :
% \begin{equation}
%   \label{eq:21}
%   S_c = \int \frac{d^2 x}{a^2} \Bigl( \gamma (\bfn \cdot \bft_3)^2 + \lambda \bigl[ (\bfn \cdot \bft_2)^3 - 3 (\bfn \cdot \bft_2) (\bfn \cdot \bft_1)^2 \bigr]^2 \Bigr) \,.
% \end{equation}
% Choosing $\bft_a = \bfe_a$, where $\bfe_a^i = \delta_{ia}$, and writing the unit vector $\bfn = \bigl( \sin \theta \cos \phi, \sin \theta \sin \phi, \cos \theta  \bigr)$, the coupling reads $S_c = \int d^2x \bigl[ \gamma \cos^2 \theta  + \lambda \sin^6 (\theta) \sin^2(3 \phi) \bigr]/2a^2$ 
% \begin{align}
%   \label{eq:3}
%   S_c = \frac12 \int d^2x \Bigl[ \frac{\gamma}{a^2} \cos^2 \theta  + \frac{\lambda}{a^2} \sin^6 (\theta) \sin^2(3 \phi) \Bigr] .
% \end{align}
A positive $\gamma>0$ favors coplanarity, whereas $\gamma<0$ induces
$\bfn$ to be perpendicular to the plane of the triangular
magnetization. 
% Depending on the sign of $\gamma$ the magnetic order parameter of the
% honeycomb lattice $\bfn$ wants to either lie within the plane of the
% triangular magnetization ($\gamma > 0$) or point perpendicular to it
% ($\gamma < 0$).
The six-fold anisotropy term $\lambda$ is relevant only
for $\gamma > 0$.
% Below we demonstrate that a Holstein-Primakoff
% $1/S$-expansion around the classical ground state
% yields $\gamma \simeq \mathcal{O}(J_{th}^2)$ and
% $\lambda \simeq \mathcal{O}(J_{th}^6) \ll \gamma$.

Heuristically, we expect $\gamma>0$, favoring
coplanarity: spins on the honeycomb lattice 
can minimize their energy by aligning
themselves perpendicular to the fluctuation
Weiss field from the triangular lattice~\cite{PhysRevLett.62.2056}.
To confirm this reasoning, 
% To analyze the quantum and thermal fluctuations of the coupling
% constants $\gamma $ and $\lambda $,  
we have performed a Holstein-Primakov spin
wave analysis of Eq.~\eqref{eq:1}. Our results for the fluctuation
correction to the free energy for arbitrary angles $\alpha $ and
$\beta $ between the two sublattices are given in Fig.~\ref{fig:1}(c-d)
and show that $\gamma >0$. 
% Similar to the tendency in a uniaxial
% system that favors colinearity, a biaxial system prefers a coplanar
% state~\cite{PhysRevLett.68.855}. 
For small $J_{th}$ we find $\gamma =
(J_{th}/\bar{J})^2 A_{\gamma }\left(J_{tt}/J_{hh}, \bar{J}/T\right) $
is the dominant term in the potential, while
$\lambda =(J_{th}/\bar{J})^6 A_{\lambda }\left(J_{tt}/J_{hh},
  \bar{J}/T \right) $, where $\bar{J} = \sqrt{J_{tt}J_{hh}}$ and the
$A_{\gamma ,\lambda }$ are functions that depend weakly on
$J_{tt}/J_{hh}$. 
% 
% The dominant term $\gamma \cos ^{2}\beta $ forces the
% direction of the uniaxial order parameter of the honeycomb lattice to
% be in the plane of the order parameter of the triangular lattice.

% A Holstein-Primakoff $1/S$-expansion around the classical ground state yields $\gamma \simeq \mathcal{O}(J_{th}^2)$ and $\lambda \simeq \mathcal{O}(J_{th}^6) \ll \gamma$. 
% Specifically, we find that the angle dependent free energy change due to non-zero $J_{th}$ can be written as $\delta F(J_{th} , T, \theta, \phi) = \zeta_Q(\theta,\phi) J_{hh} + \zeta_T(\theta, \phi) T$~\cite{PhysRevLett.64.88} with quantum and thermal contribution ($\nu = Q, T$)
% \begin{equation}
%   \label{eq:32}
%   \zeta_\nu =   \mathcal{A}_\nu j_{th}^2 f(j_{tt}) \cos^2 \theta + \mathcal{B}_\nu j_{th}^6 g(j_{tt}) \sin^6 \theta \sin^2 (3 \phi) \,,
% \end{equation}
% with $j_{th} = J_{th}/J_{hh}$, $j_{tt} = J_{tt}/J_{hh}$, $f(x) = (1 + x)/2x$ and $g(x) = ...$. The numerical coefficients read $\mathcal{A}_T = 0.48$, $\mathcal{B}_{T} = 3 \times 10^{-3}$, $\mathcal{A}_{Q} = 0.022$ and $\mathcal{B}_{Q} = 1.3 \times 10^{-4}$. This gives rise to the potential term $S_c =\frac12 \int d^2x (\zeta_Q J_{hh} + \zeta_T T)/T a^2$ in the action in Eq.~\eqref{eq:2}.

As temperature is reduced, 
the two sublattices enter
a \emph{coplanar regime}. 
% due to the dominant potential term $\propto
% \gamma (\bfn \cdot \bft_3)^2$,. 
The temperature scale for this
crossover is easily determined from standard scaling arguments and
yields the coplanar crossover temperature (see Fig.~\ref{fig:3}(a))
\begin{equation}
  \label{eq:23}
  T_{\text{cp}} \simeq \frac{J_{hh}S^2}{1 + \ln (1/\gamma)/4 \pi}
\end{equation}
in case where $J_{hh} < J_{tt}$. In the opposite regime $J_{tt} < J_{hh}$ we obtain an implicit expression for $T_{\text{cp}}$ that also approaches zero only logarithmically for $\gamma \rightarrow 0$. The crossover temperature in Eq.~\eqref{eq:23} follows from the known flow equation $\frac{d}{dl} K = - 1/2\pi$ for the stiffness $K$ with running cutoff $\Lambda(l) = a_0^{-1} e^{-l}$ and the flow of the coplanar potential amplitude $\gamma(l) = \gamma \exp( 2 l)$ that is determined by its engineering dimension. While the spin stiffnesses are reduced at longer length-scales, the potential term grows, and scaling stops when $\gamma(l_\gamma) = 1$, which defines a length-scale $a_\gamma = a_0 e^{l_\gamma} \simeq a_0(J_{hh}/J_{th})^2$. The coplanar crossover takes place when this length-scale is comparable to the shorter of the magnetic correlation lengths on the two sublattices. From the known flow equation of the $O(3)/O(2)$ and the $SO(3)$ NLSM further follows that the stiffnesses of the triangular lattice approach an isotropic fixed point~\cite{PhysRevLett.64.3175}. The six-fold symmetric potential $\propto \lambda$ flows to larger values, yet due to $\lambda \ll \gamma$ holds that $\lambda(l_\gamma) \simeq \mathcal{O}(J_{th}^4) \ll 1$. 

% We determine the sign and magnitude of the coupling coefficients $\gamma$ and $\lambda$ by performing a Holstein-Primakoff $1/S$-expansion of the Hamiltonian in Eq.~\eqref{eq:1}. After an Bogoliubov transformation it takes the form 
% \begin{align}
%   \label{eq:33}
%   H &=  - 3 N_L  \bigl(J_{hh} + \frac{J_{tt}}{2}  \bigr) (S^2 + S) \nonumber \\ &+ \sum_{\bfp \in \text{MBZ}} \sum_{\alpha = 1}^9 E_{\alpha, \bfp} \bigl( b^\dag_{\alpha,\bfp} b_{\alpha, \bfp} + \frac12  \bigr) + \mathcal{O}(1/S) \,,
% \end{align}
% where the spin-wave spectrum $E_{\alpha, \bfp}(\theta, \phi)$ is obtained numerically. The angle dependent free energy change due to non-zero $J_{th}$ can be written as $\delta F(J_{th} , T, \theta, \phi) = \zeta_Q(\theta,\phi) J_{hh} + \zeta_T(\theta, \phi) T$~\cite{PhysRevLett.64.88} with quantum and thermal contribution ($\nu = Q, T$)
% \begin{equation}
%   \label{eq:32}
%   \zeta_\nu =   \mathcal{A}_\nu j_{th}^2 f(z) \cos^2 \theta + \mathcal{B}_\nu j_{th}^6 g(z) \sin^6 \theta \sin^2 (3 \phi) \,,
% \end{equation}
% with $j_{th} = J_{th}/J_{hh}$, $z = J_{hh}/J_{tt}$, $f(x) = (1 + x)/2$ and $g(x) = ...$. \PPO{find $g(x)$} The numerical coefficients read $\mathcal{A}_T = 0.48$, $\mathcal{B}_{T} = 3 \times 10^{-3}$, $\mathcal{A}_{Q} = 0.022$ and $\mathcal{B}_{Q} = 1.3 \times 10^{-4}$. The gives rise to the potential term $S_c =\frac12 \int d^2x (\zeta_Q J_{hh} + \zeta_T T)/T a^2$ in the action in Eq.~\eqref{eq:2}.

% \emph{Coplanar regime action.--}

Once the two sublattices  are coplanar,
% lie in the 
% have been forced to fluctuate locally in the
% same plane, 
their dynamics are intimately connected. To describe
this regime we impose a hard-core constraint: $\bfn
\perp \bft_3$, \emph{i.e.}, $\beta = \pi/2$. It is now convenient to introduce a second triad
$\bfh_{1,2,3}$, defining an $SO(3)$ matrix $h = \bigl( \bfh_1, \bfh_2, \bfh_3
\bigr)$ that describes the magnetic order on the honeycomb lattice
with $\bfh_1= \bfn$. The coplanar constraint is expressed as $t = h U$
where $U = \exp( i \alpha \tau_3 )$  determines the relative inplane
orientation of the two sublattices, defined by the angle $\alpha $. 
We describe $h$ in terms of three Euler angles, 
$h
= e^{-i \phi \tau_3} e^{-i \theta \tau_1} e^{-i \psi \tau_3}$. 
Here, the $\tau_{a}$ satisfy the $SU(2)$ algebra $\bigl[ \tau_a,
\tau_b \bigr] = i \epsilon_{abc} \tau_c$ 
and take the adjoint form $(\tau_a)_{bc}
= i \epsilon_{bac}$. 
% The degrees of freedom of 
The coplanar system is thus  determined
by an $SO(3)\times U (1)$ order parameter, defined by
three Euler angles and a single relative phase $\alpha $.

% In
% addition, the relative orientation of the two sublattices is
% characterized by the $U(1)$ angle $\alpha$.

To analyze this coupled
problem we write the action in the form $S=S_{X}+S_{c}$, where
\begin{equation}
  \label{eq:37}
  S_{X}=\frac{1}{2}\int d^{2}xg_{ij}[ X( x) ] \partial _{\mu
  }X^{i}( x ) \partial _{\mu }X^{j}( x ) 
\end{equation}
with coordinates $X=\left( \phi ,\theta ,\psi ,\alpha \right) $ and stiffness tensor 
\begin{equation}
  \label{eq:36}
  g= \begin{pmatrix} g^{SO\left( 3\right) } & \mathcal{K}^{T} \\ 
    \mathcal{K} & I_{\alpha }
  \end{pmatrix},
\end{equation}
where 
\begin{equation*}
  \label{eq:34}
  g^{SO\left( 3\right) }_{ij}= \begin{pmatrix} 
    (I_{1}s_\psi^2 +I_{2} c_\psi^2) s_\theta^2 +I_{3} c_\theta^2  & (I_{1}-I_{2}) c_\psi s_\theta s_\psi  & I_{3} c_\theta  \\ 
    (I_{1} - I_{2} ) c_\psi s_\theta s_\psi \  & I_{1} c_\psi^2 +I_{2} s_\psi^2  & 0 \\ 
    I_{3} c_\theta  & 0 & I_{3}   \end{pmatrix} ,
\end{equation*}
% \begin{equation}
%   \label{eq:25}
%   S = \frac12 \int d^2 x \, g_{ij}[\boldsymbol{\omega}(x)] \partial_\mu \omega^i(x) \partial_\mu \omega^j(x) + S_c 
% \end{equation}
% with contravariant coordinates $\boldsymbol{\omega} = (\phi, \theta, \psi, \alpha)$ and covariant metric tensor
% \begin{equation}
%   \label{eq:26}
%   g = \begin{pmatrix} g^{O(3)} & c^T \\ c & I_\alpha \end{pmatrix} \,.
% \end{equation}
% Here, $g^{O(3)}_{ij}$ is the usual metric tensor of the $O(3)$ problem, which reads explicitly
% \begin{equation}
%   \label{eq:27}
%   g^{O(3)} = \begin{pmatrix} 
%     (I_{1}s_\psi^2 +I_{2} c_\psi^2) s_\theta^2 +I_{3} c_\theta^2  & (I_{1}-I_{2}) c_\psi s_\theta s_\psi  & I_{3} c_\theta  \\ 
%     (I_{1} - I_{2} ) c_\psi s_\theta s_\psi \  & I_{1} c_\psi^2 +I_{2} s_\psi^2  & 0 \\ 
%     I_{3} c_\theta  & 0 & I_{3}   \end{pmatrix} \,,
% \end{equation}
with $s_{X^j} = \sin X^j$ and $c_{X^j} = \cos X^j$. In our system we
find $I_1 = K_1 + K_3$, $I_2 = K_1 + K_3 + K$, $I_3 = 2 K_1 + K$,
which are set by the stiffnesses of the two sublattices at
$l=l_\gamma$. The $U(1)$ degree of freedom $\alpha$ has an
initial stiffness $I_\alpha = 2 K_1(l_\gamma)$ and is 
coupled to the non-Abelian $SO(3)$ sector 
by the 
term $\mathcal{K} = \frac{\kappa}{2} ( c_\theta, 0,
1)$ in the four-dimensional metric, where $\kappa = 4
K_1(l_\gamma)$. The six-fold potential $S_c(\beta=\frac{\pi}{2} ) = \frac12
\lambda \int d^2x \sin^2 (3 \alpha)$ is a small but relevant
perturbation to $S_X$. 
At length-scales where $\lambda$ is small, the anisotropy 
$S_c$ and the gradient term $S_{X}$~\eqref{eq:37} 
is the action of a classical string in a four dimensional space with
coordinates $X\left( x\right) $ at the two dimensional worldsheet
point $x$, with metric tensor $g_{ij}[X]$. Under 
coordinate transformations 
$X_{i}\rightarrow X^{\prime }_{i} $, $S_{X}$ in
Eq.~\eqref{eq:37} is invariant, 
with
transformed metric $g_{lm}^{\prime }=g_{ij}\frac{\partial
  X^{i}}{\partial X^{\prime l}}\frac{\partial X^{j}}{\partial X^{\prime
    m}}$. 
Like Einstein's theory of gravity, 
this covariance tells us that the long-wavelength action $S_{X}$
is co-ordinate independent
and only depends
on the geometric aspects of the mapping
$X\left(x\right) $ of the wordsheet to the compact four-dimensional
space of the coordinate $X$. 
The renormalization group (RG) flow
of the metric tensor must also be covariant under co-ordinate
transformations, and following
the geometric
formulation of the NLSM by Friedan~\cite{PhysRevLett.45.1057}, to two
loop order takes
the form
% we can obtain the renormalization group flow equation of the above
% action from Friedan's geometric formulation of the
% NLSM~\cite{PhysRevLett.45.1057}.
\begin{equation}
  \label{eq:28}
  \frac{d g_{ij}}{dl} = \frac{1}{2 \pi} R_{ij} - \frac{1}{8 \pi^2} R_i{}^{klm} R_{jklm} \,,
\end{equation}
where $R^{iklm}$ is the Riemann curvature tensor and $R_{ij} =
R^k{}_{ikj}$ is the Ricci tensor~\cite{OneLoopPolykovRG}. This expression defines a generalized
Ricci flow~\cite{Hamilton_RicciFlow_1982}.
The Riemann tensor is determined by the Christoffel symbols $\Gamma^i_{jk}
= \frac12 g^{il} (g_{jl,k} + g_{kl,j} - g_{jk,l})$ as $R^k{}_{lij} =
\Gamma^k_{lj,i} - \Gamma^k_{li,j} + \Gamma^k_{ni} \Gamma^n_{lj} -
\Gamma^k_{nj} \Gamma^n_{li}$. The flow equations of our five coupling
constants $I_j$, $I_\alpha$ and $\kappa$ follow from
Eq.~\eqref{eq:28}.

% \emph{Low-temperature phase diagram.--} 

A key insight into the low energy phase diagram is obtained by noting
the coupling term $\mathcal{K}$ can  be
eliminated via a coordinate transformation $\psi \rightarrow \psi' =
\psi + r \alpha$ with $r = \kappa/2 I_3$. This yields a metric $g$ in
Eq.~\eqref{eq:36} with $\mathcal{K} = 0$, $I_\alpha \rightarrow
I'_\alpha = I_\alpha - \kappa^2/4 I_3$ yet with $g^{SO(3)}$ that
depends on the $U(1)$ phase $\alpha$ via the above shift of the Euler
angle $\psi$. This gauge transformation to the appropriate center of
mass coordinates allows for clear criteria when the $U(1)$ sector of
the theory decouples from the $SO(3)$ sector: if either $|I_1 - I_2|
\ll \sqrt{I_1 I_2}$ or $r \ll 1$ it follows that $g^{SO(3)}$ becomes
independent of $\alpha$ and the $U(1)$ phase decouples from the
dynamics of the non-collinear magnetic degrees of freedom. The first
criterion follows from the fact that $g^{SO(3)}$ is independent of
$\psi$ if $I_1 = I_2$, while the second criterion implies that the
shift in $\psi$ is negligible. From Eq.~\eqref{eq:28} follows after a
lengthy but straightforward calculation that $I_{1,2,3}$ flow to an
isotropic fixed point, while the dimensionless variable $r$ follows
the flow equation (for simplicity we only list the one loop result,
the two loop correction does not change our conclusions):
\begin{equation}
  \label{eq:29}
  \frac{dr}{dl} = - r \frac{(I_1 - I_2)^2}{4 \pi I_1 I_2 I_3} \,.
\end{equation}
Thus, if the initial anisotropy $|I_1 - I_2| = K$ is weak, which happens for $J_{hh} \ll J_{tt}$, the coupling $r$ does not change much. The $SO(3)$ sector, however, quickly becomes isotropic in the $1$-$2$--plane leading to a decoupling of the $U(1)$ phase. On the other hand, in the limit of strong anisotropy for $J_{hh} \gg J_{tt}$, where $|I_{1} - I_2|$ is not small, we find that $r$ vanishes rapidly. In both cases follows that the phase angle $\alpha$ emerges as an independent degree of freedom. The $\beta$-function for the reduced phase stiffness $I'_\alpha = I_\alpha - \kappa^2/4 I_3$ follows from Eq.~\eqref{eq:28} as
\begin{equation}
  \label{eq:30}
  \frac{d I'_\alpha}{dl} = \beta_\alpha = \frac{(I_1 - I_2)^2 r^2}{4 \pi I_1 I_2} \,,
\end{equation}
and does, as expected, approach zero once either of the two decoupling conditions are fulfilled. Thus, perturbatively no renormalization of the stiffness $I'_\alpha$ takes place. In Fig.~\ref{fig:3}(b) we present the coplanar renormalization group flow for two different sets of parameters corresponding to weak and strong initial anisotropy.
% The metric becomes completely flat in this sector. 
An interesting aspect of the decoupling follows from the Ricci scalar $R=g^{ij}R_{ji}$:
\begin{equation}
  \label{eq:38}
  R=R^{SO\left( 3\right) }-\frac{1}{2\pi I'_{\alpha}} \beta _{\alpha}
\end{equation}
where $R^{SO\left( 3\right) }=\sum_{j=1}^{3}\left( I_{j}^{-1}-\frac{1}{2I_{1}I_{2}I_{3}}I_{j}^{2}\right) $ is the Ricci scalar of the $SO\left(3\right) $ sector. Once the decoupling takes place, $\beta _{\alpha} \rightarrow 0$ and the $U\left( 1\right) $ sector becomes flat. On the other hand  $R\rightarrow R^{SO\left( 3\right) }$ grows under renormalization since the stiffnesses $I_{j}$ decrease. Thus, we arrive at a flat one dimensional sector weakly coupled to an three-dimensional manifold with large curvature. This "curling-up" and asymptotic decoupling of a subspace  may serve as a toy model for compactification.

Since the decoupling emerges rapidly in both limits $J_{hh} \ll J_{tt} $ and $J_{hh} \gg J_{tt}$, we find that $\lambda$, whose flow is governed by $\frac{d}{dl} \lambda = ( 2 - 9/\pi I'_\alpha) \lambda$, is still small at the decoupling lengthscale. The resulting low-energy theory corresponds to $S = S_{SO(3)} + S_{\mathbb{Z}_6}$ with 
\begin{equation}
  \label{eq:31}
  S_{\mathbb{Z}_6} = \frac12 \int d^2x \bigl[ (I'_\alpha (\partial_\mu \alpha)^2 + \lambda \sin^2 (3 \alpha) \bigr] \,.
\end{equation}
This is the well-known six state clock model that exhibits two
consecutive BKT transitions~\cite{PhysRevB.16.1217}: one at
$T_{\text{BKT}}^>$ that separates a high temperature disordered phase
from a low temperature critical phase, where correlations $\av{\exp[i
  (\alpha(x) - \alpha(x'))]}$ decay with a power-law in $|x -
x'|$; a second at $T_{\text{BKT}}^<$
where the $\mathbb{Z}_6$ symmetry is spontaneously broken, leading to true
long-range order with $\alpha=n \pi/3$ ($n \in \{ 1, \ldots, 6\}$). It is crucial that
the decoupling of the $U(1)$ phase occurs first, otherwise 
the $SO(3)$ sector would screen the long-range interactions 
between topological defects -- vortices at $T_{\text{BKT}}^{>}$ or
domain walls at $T_{\text{BKT}}^<$ that are responsible for the BKT
transitions and 
the intermediate critical phase.

Following the RG program of the BKT problem for Eq.~\eqref{eq:31} we need to take into account that the size of the vortex is now given by the coplanar lengthscale $a_\gamma \gg a_0$~\cite{ChaikinLubensky-Book,PhysRevLett.109.155703}. We determine the vortex unbinding transition temperature $T_{\text{BKT}}^>$ implicitly via
\begin{equation}
  \label{eq:22}
  I'_\alpha(T_{\text{BKT}}^>)^{-1} = \frac{\pi}{2 + 4 \pi y(T_{\text{BKT}}^>) }
\end{equation}
with fugacity $y = e^{-S_c} a_\gamma^2/a_0^2$ and core action $S_c
\simeq \pi \{1+ \text{min}(K,K_t)\} $. From Eq.~\eqref{eq:22} we predict that
$T_{\text{BKT}}^> \lesssim T_{\text{cp}}$, \emph{i.e.}, the BKT
transition is only numerically smaller than the coplanar crossover
temperature. The system enters the critical phase soon after it
becomes coplanar. Similarly, it follows from
Ref.~\cite{PhysRevB.16.1217} that $T_{\text{BKT}}^<$ and
$T_{\text{BKT}}^>$ are of the same order of magnitude. The resulting phase diagram is shown in Fig.~\ref{fig:3}(a).  
% We note that the relative order of spins effects a distortion of the
% lattice. The corresponding elasticity modulus is thus strictly zero
% in the critical phase (``lattice softening'') where the correlation
% length is infinite~\cite{PhysRevLett.105.157003}.

%%% Joerg and Peter - we ran out of energy and time to work on the final
%% paragraph. This one is basically just a place holder.
In summary we have presented a 2D Heisenberg model on a decorated triangular lattice where short wavelength thermal fluctuations select long-range $\mathbb{Z}_6$ order. This is preceded in temperature by an emergent critical phase that is framed by two BKT transitions.
We have written the action of this model as a classical 4D string theory where the spin stiffness is determined by the 
metric tensor of the manifold; the scaling equations are then extracted as components of the resulting Ricci flow.  We note that the decoupling
of the $U(1)$ degree of freedom corresponds to a dimensional reduction of the analogous string theory and thus to a toy
model of compactification.  Finally we note that the emergence of massless modes in collective mode massive theories could have interesting implications for two-dimensional field theories.

\begin{acknowledgments}
  We acknowledge useful discussions with S. T. Carr, R. Fernandes,
  E. J. K\"onig, D. Nelson, V. Oganesyan, P. Ostrovsky, N. Perkins, J. Reuther,
  S. Sondhi, and O. Sushkov. The Young Investigator Group of P.P.O. received financial support from the ``Concept for the Future'' of the KIT within the framework of the German Excellence Initiative. This work was supported by DOE grant DE-FG02-99ER45790 (P. Coleman) and SEPNET (P.C., P.C. and J.S.). P.C., P.C. and J.S. acknowledge the hospitality of Royal Holloway, University of London where this work was begun.
\end{acknowledgments}

\emph{Note added}.-- After obtaining these results we learned
of two recent studies: one on a Kitaev-Heisenberg
model, where an emergent $\mathbb{Z}_6$-symmetry results from a
conceptually different mechanism~\cite{PhysRevLett.109.187201}, a
second on itinerant systems where an emergent $\mathbb{Z}_4$ Potts model
appears~\cite{PhysRevB.86.115443}.

%\bibliography{Biblio}

\begin{thebibliography}{36}%
\makeatletter
\providecommand \@ifxundefined [1]{%
 \@ifx{#1\undefined}
}%
\providecommand \@ifnum [1]{%
 \ifnum #1\expandafter \@firstoftwo
 \else \expandafter \@secondoftwo
 \fi
}%
\providecommand \@ifx [1]{%
 \ifx #1\expandafter \@firstoftwo
 \else \expandafter \@secondoftwo
 \fi
}%
\providecommand \natexlab [1]{#1}%
\providecommand \enquote  [1]{``#1''}%
\providecommand \bibnamefont  [1]{#1}%
\providecommand \bibfnamefont [1]{#1}%
\providecommand \citenamefont [1]{#1}%
\providecommand \href@noop [0]{\@secondoftwo}%
\providecommand \href [0]{\begingroup \@sanitize@url \@href}%
\providecommand \@href[1]{\@@startlink{#1}\@@href}%
\providecommand \@@href[1]{\endgroup#1\@@endlink}%
\providecommand \@sanitize@url [0]{\catcode `\\12\catcode `\$12\catcode
  `\&12\catcode `\#12\catcode `\^12\catcode `\_12\catcode `\%12\relax}%
\providecommand \@@startlink[1]{}%
\providecommand \@@endlink[0]{}%
\providecommand \url  [0]{\begingroup\@sanitize@url \@url }%
\providecommand \@url [1]{\endgroup\@href {#1}{\urlprefix }}%
\providecommand \urlprefix  [0]{URL }%
\providecommand \Eprint [0]{\href }%
\providecommand \doibase [0]{http://dx.doi.org/}%
\providecommand \selectlanguage [0]{\@gobble}%
\providecommand \bibinfo  [0]{\@secondoftwo}%
\providecommand \bibfield  [0]{\@secondoftwo}%
\providecommand \translation [1]{[#1]}%
\providecommand \BibitemOpen [0]{}%
\providecommand \bibitemStop [0]{}%
\providecommand \bibitemNoStop [0]{.\EOS\space}%
\providecommand \EOS [0]{\spacefactor3000\relax}%
\providecommand \BibitemShut  [1]{\csname bibitem#1\endcsname}%
\let\auto@bib@innerbib\@empty
%</preamble>
\bibitem [{\citenamefont {{H}ohenberg}(1967)}]{PhysRev.158.383}%
  \BibitemOpen
  \bibfield  {author} {\bibinfo {author} {\bibfnamefont {P.~C.}\ \bibnamefont
  {{H}ohenberg}},\ }\href {\doibase 10.1103/PhysRev.158.383} {\bibfield
  {journal} {\bibinfo  {journal} {Phys. Rev.}\ }\textbf {\bibinfo {volume}
  {158}},\ \bibinfo {pages} {383} (\bibinfo {year} {1967})}\BibitemShut
  {NoStop}%
\bibitem [{\citenamefont {{M}ermin}\ and\ \citenamefont
  {{W}agner}(1966)}]{mermin_absence_1966}%
  \BibitemOpen
  \bibfield  {author} {\bibinfo {author} {\bibfnamefont {N.~D.}\ \bibnamefont
  {{M}ermin}}\ and\ \bibinfo {author} {\bibfnamefont {H.}~\bibnamefont
  {{W}agner}},\ }\href {\doibase 10.1103/PhysRevLett.17.1307} {\bibfield
  {journal} {\bibinfo  {journal} {Phys. Rev. Lett.}\ }\textbf {\bibinfo
  {volume} {17}},\ \bibinfo {pages} {1307} (\bibinfo {year}
  {1966})}\BibitemShut {NoStop}%
\bibitem [{\citenamefont {{C}handra}\ \emph {et~al.}(1990)\citenamefont
  {{C}handra}, \citenamefont {{C}oleman},\ and\ \citenamefont
  {{L}arkin}}]{PhysRevLett.64.88}%
  \BibitemOpen
  \bibfield  {author} {\bibinfo {author} {\bibfnamefont {P.}~\bibnamefont
  {{C}handra}}, \bibinfo {author} {\bibfnamefont {P.}~\bibnamefont
  {{C}oleman}}, \ and\ \bibinfo {author} {\bibfnamefont {A.~I.}\ \bibnamefont
  {{L}arkin}},\ }\href {\doibase 10.1103/PhysRevLett.64.88} {\bibfield
  {journal} {\bibinfo  {journal} {Phys. Rev. Lett.}\ }\textbf {\bibinfo
  {volume} {64}},\ \bibinfo {pages} {88} (\bibinfo {year} {1990})}\BibitemShut
  {NoStop}%
\bibitem [{\citenamefont {{C}apriotti}\ \emph {et~al.}(2004)\citenamefont
  {{C}apriotti}, \citenamefont {{F}ubini}, \citenamefont {{R}oscilde},\ and\
  \citenamefont {{T}ognetti}}]{PhysRevLett.92.157202}%
  \BibitemOpen
  \bibfield  {author} {\bibinfo {author} {\bibfnamefont {L.}~\bibnamefont
  {{C}apriotti}}, \bibinfo {author} {\bibfnamefont {A.}~\bibnamefont
  {{F}ubini}}, \bibinfo {author} {\bibfnamefont {T.}~\bibnamefont
  {{R}oscilde}}, \ and\ \bibinfo {author} {\bibfnamefont {V.}~\bibnamefont
  {{T}ognetti}},\ }\href {\doibase 10.1103/PhysRevLett.92.157202} {\bibfield
  {journal} {\bibinfo  {journal} {Phys. Rev. Lett.}\ }\textbf {\bibinfo
  {volume} {92}},\ \bibinfo {pages} {157202} (\bibinfo {year}
  {2004})}\BibitemShut {NoStop}%
\bibitem [{\citenamefont {{W}eber}\ \emph {et~al.}(2003)\citenamefont
  {{W}eber}, \citenamefont {{C}apriotti}, \citenamefont {{M}isguich},
  \citenamefont {{B}ecca}, \citenamefont {{E}lhajal},\ and\ \citenamefont
  {{M}ila}}]{PhysRevLett.91.177202}%
  \BibitemOpen
  \bibfield  {author} {\bibinfo {author} {\bibfnamefont {C.}~\bibnamefont
  {{W}eber}}, \bibinfo {author} {\bibfnamefont {L.}~\bibnamefont
  {{C}apriotti}}, \bibinfo {author} {\bibfnamefont {G.}~\bibnamefont
  {{M}isguich}}, \bibinfo {author} {\bibfnamefont {F.}~\bibnamefont {{B}ecca}},
  \bibinfo {author} {\bibfnamefont {M.}~\bibnamefont {{E}lhajal}}, \ and\
  \bibinfo {author} {\bibfnamefont {F.}~\bibnamefont {{M}ila}},\ }\href
  {\doibase 10.1103/PhysRevLett.91.177202} {\bibfield  {journal} {\bibinfo
  {journal} {Phys. Rev. Lett.}\ }\textbf {\bibinfo {volume} {91}},\ \bibinfo
  {pages} {177202} (\bibinfo {year} {2003})}\BibitemShut {NoStop}%
\bibitem [{\citenamefont {{M}ulder}\ \emph {et~al.}(2010)\citenamefont
  {{M}ulder}, \citenamefont {{G}anesh}, \citenamefont {{C}apriotti},\ and\
  \citenamefont {{P}aramekanti}}]{PhysRevB.81.214419}%
  \BibitemOpen
  \bibfield  {author} {\bibinfo {author} {\bibfnamefont {A.}~\bibnamefont
  {{M}ulder}}, \bibinfo {author} {\bibfnamefont {R.}~\bibnamefont {{G}anesh}},
  \bibinfo {author} {\bibfnamefont {L.}~\bibnamefont {{C}apriotti}}, \ and\
  \bibinfo {author} {\bibfnamefont {A.}~\bibnamefont {{P}aramekanti}},\ }\href
  {\doibase 10.1103/PhysRevB.81.214419} {\bibfield  {journal} {\bibinfo
  {journal} {Phys. Rev. B}\ }\textbf {\bibinfo {volume} {81}},\ \bibinfo
  {pages} {214419} (\bibinfo {year} {2010})}\BibitemShut {NoStop}%
\bibitem [{\citenamefont {{H}enley}(2010)}]{annurev-conmatphys-070909-104138}%
  \BibitemOpen
  \bibfield  {author} {\bibinfo {author} {\bibfnamefont {C.~L.}\ \bibnamefont
  {{H}enley}},\ }\href@noop {} {\bibfield  {journal} {\bibinfo  {journal}
  {Annu. Rev. Cond. Mat. Phys.}\ }\textbf {\bibinfo {volume} {1}},\ \bibinfo
  {pages} {179} (\bibinfo {year} {2010})}\BibitemShut {NoStop}%
\bibitem [{\citenamefont {{Z}hitomirsky}(2008)}]{PhysRevB.78.094423}%
  \BibitemOpen
  \bibfield  {author} {\bibinfo {author} {\bibfnamefont {M.~E.}\ \bibnamefont
  {{Z}hitomirsky}},\ }\href {\doibase 10.1103/PhysRevB.78.094423} {\bibfield
  {journal} {\bibinfo  {journal} {Phys. Rev. B}\ }\textbf {\bibinfo {volume}
  {78}},\ \bibinfo {pages} {094423} (\bibinfo {year} {2008})}\BibitemShut
  {NoStop}%
\bibitem [{\citenamefont {{C}hern}\ and\ \citenamefont
  {{M}oessner}(2012)}]{arXiv:1207.4752v1}%
  \BibitemOpen
  \bibfield  {author} {\bibinfo {author} {\bibfnamefont {G.-W.}\ \bibnamefont
  {{C}hern}}\ and\ \bibinfo {author} {\bibfnamefont {R.}~\bibnamefont
  {{M}oessner}},\ }\href@noop {} {\bibfield  {journal} {\bibinfo  {journal}
  {arXiv:1207.4752v1 [cond-mat.str-el]}\ } (\bibinfo {year}
  {2012})}\BibitemShut {NoStop}%
\bibitem [{\citenamefont {{F}ernandes}\ \emph {et~al.}(2010)\citenamefont
  {{F}ernandes}, \citenamefont {{V}an{B}ebber}, \citenamefont {{B}hattacharya},
  \citenamefont {{C}handra}, \citenamefont {{K}eppens}, \citenamefont
  {{M}andrus}, \citenamefont {{M}c{G}uire}, \citenamefont {{S}ales},
  \citenamefont {{S}efat},\ and\ \citenamefont
  {{S}chmalian}}]{PhysRevLett.105.157003}%
  \BibitemOpen
  \bibfield  {author} {\bibinfo {author} {\bibfnamefont {R.~M.}\ \bibnamefont
  {{F}ernandes}}, \bibinfo {author} {\bibfnamefont {L.~H.}\ \bibnamefont
  {{V}an{B}ebber}}, \bibinfo {author} {\bibfnamefont {S.}~\bibnamefont
  {{B}hattacharya}}, \bibinfo {author} {\bibfnamefont {P.}~\bibnamefont
  {{C}handra}}, \bibinfo {author} {\bibfnamefont {V.}~\bibnamefont
  {{K}eppens}}, \bibinfo {author} {\bibfnamefont {D.}~\bibnamefont
  {{M}andrus}}, \bibinfo {author} {\bibfnamefont {M.~A.}\ \bibnamefont
  {{M}c{G}uire}}, \bibinfo {author} {\bibfnamefont {B.~C.}\ \bibnamefont
  {{S}ales}}, \bibinfo {author} {\bibfnamefont {A.~S.}\ \bibnamefont
  {{S}efat}}, \ and\ \bibinfo {author} {\bibfnamefont {J.}~\bibnamefont
  {{S}chmalian}},\ }\href {\doibase 10.1103/PhysRevLett.105.157003} {\bibfield
  {journal} {\bibinfo  {journal} {Phys. Rev. Lett.}\ }\textbf {\bibinfo
  {volume} {105}},\ \bibinfo {pages} {157003} (\bibinfo {year}
  {2010})}\BibitemShut {NoStop}%
\bibitem [{\citenamefont {{J}os\'e}\ \emph {et~al.}(1977)\citenamefont
  {{J}os\'e}, \citenamefont {{K}adanoff}, \citenamefont {{K}irkpatrick},\ and\
  \citenamefont {{N}elson}}]{PhysRevB.16.1217}%
  \BibitemOpen
  \bibfield  {author} {\bibinfo {author} {\bibfnamefont {J.~V.}\ \bibnamefont
  {{J}os\'e}}, \bibinfo {author} {\bibfnamefont {L.~P.}\ \bibnamefont
  {{K}adanoff}}, \bibinfo {author} {\bibfnamefont {S.}~\bibnamefont
  {{K}irkpatrick}}, \ and\ \bibinfo {author} {\bibfnamefont {D.~R.}\
  \bibnamefont {{N}elson}},\ }\href {\doibase 10.1103/PhysRevB.16.1217}
  {\bibfield  {journal} {\bibinfo  {journal} {Phys. Rev. B}\ }\textbf {\bibinfo
  {volume} {16}},\ \bibinfo {pages} {1217} (\bibinfo {year}
  {1977})}\BibitemShut {NoStop}%
\bibitem [{\citenamefont {{O}rtiz}\ \emph {et~al.}(2012)\citenamefont
  {{O}rtiz}, \citenamefont {{C}obanera},\ and\ \citenamefont
  {{N}ussinov}}]{Ortiz2012780}%
  \BibitemOpen
  \bibfield  {author} {\bibinfo {author} {\bibfnamefont {G.}~\bibnamefont
  {{O}rtiz}}, \bibinfo {author} {\bibfnamefont {E.}~\bibnamefont {{C}obanera}},
  \ and\ \bibinfo {author} {\bibfnamefont {Z.}~\bibnamefont {{N}ussinov}},\
  }\href {\doibase 10.1016/j.nuclphysb.2011.09.012} {\bibfield  {journal}
  {\bibinfo  {journal} {Nucl. Phys. B}\ }\textbf {\bibinfo {volume} {854}},\
  \bibinfo {pages} {780} (\bibinfo {year} {2012})}\BibitemShut {NoStop}%
\bibitem [{\citenamefont {{B}lankschtein}\ \emph {et~al.}(1984)\citenamefont
  {{B}lankschtein}, \citenamefont {{M}a}, \citenamefont {{B}erker},
  \citenamefont {{G}rest},\ and\ \citenamefont
  {{S}oukoulis}}]{PhysRevB.29.5250}%
  \BibitemOpen
  \bibfield  {author} {\bibinfo {author} {\bibfnamefont {D.}~\bibnamefont
  {{B}lankschtein}}, \bibinfo {author} {\bibfnamefont {M.}~\bibnamefont
  {{M}a}}, \bibinfo {author} {\bibfnamefont {A.~N.}\ \bibnamefont {{B}erker}},
  \bibinfo {author} {\bibfnamefont {G.~S.}\ \bibnamefont {{G}rest}}, \ and\
  \bibinfo {author} {\bibfnamefont {C.~M.}\ \bibnamefont {{S}oukoulis}},\
  }\href {\doibase 10.1103/PhysRevB.29.5250} {\bibfield  {journal} {\bibinfo
  {journal} {Phys. Rev. B}\ }\textbf {\bibinfo {volume} {29}},\ \bibinfo
  {pages} {5250} (\bibinfo {year} {1984})}\BibitemShut {NoStop}%
\bibitem [{\citenamefont {{I}sakov}\ and\ \citenamefont
  {{M}oessner}(2003)}]{PhysRevB.68.104409}%
  \BibitemOpen
  \bibfield  {author} {\bibinfo {author} {\bibfnamefont {S.~V.}\ \bibnamefont
  {{I}sakov}}\ and\ \bibinfo {author} {\bibfnamefont {R.}~\bibnamefont
  {{M}oessner}},\ }\href {\doibase 10.1103/PhysRevB.68.104409} {\bibfield
  {journal} {\bibinfo  {journal} {Phys. Rev. B}\ }\textbf {\bibinfo {volume}
  {68}},\ \bibinfo {pages} {104409} (\bibinfo {year} {2003})}\BibitemShut
  {NoStop}%
\bibitem [{\citenamefont {{D}uan}\ \emph {et~al.}(2003)\citenamefont {{D}uan},
  \citenamefont {{D}emler},\ and\ \citenamefont
  {{L}ukin}}]{PhysRevLett.91.090402}%
  \BibitemOpen
  \bibfield  {author} {\bibinfo {author} {\bibfnamefont {L.-M.}\ \bibnamefont
  {{D}uan}}, \bibinfo {author} {\bibfnamefont {E.}~\bibnamefont {{D}emler}}, \
  and\ \bibinfo {author} {\bibfnamefont {M.~D.}\ \bibnamefont {{L}ukin}},\
  }\href {\doibase 10.1103/PhysRevLett.91.090402} {\bibfield  {journal}
  {\bibinfo  {journal} {Phys. Rev. Lett.}\ }\textbf {\bibinfo {volume} {91}},\
  \bibinfo {pages} {090402} (\bibinfo {year} {2003})}\BibitemShut {NoStop}%
\bibitem [{\citenamefont {{L}ubasch}\ \emph {et~al.}(2011)\citenamefont
  {{L}ubasch}, \citenamefont {{M}urg}, \citenamefont {{S}chneider},
  \citenamefont {{C}irac},\ and\ \citenamefont {{B}a\
  nuls}}]{PhysRevLett.107.165301}%
  \BibitemOpen
  \bibfield  {author} {\bibinfo {author} {\bibfnamefont {M.}~\bibnamefont
  {{L}ubasch}}, \bibinfo {author} {\bibfnamefont {V.}~\bibnamefont {{M}urg}},
  \bibinfo {author} {\bibfnamefont {U.}~\bibnamefont {{S}chneider}}, \bibinfo
  {author} {\bibfnamefont {J.~I.}\ \bibnamefont {{C}irac}}, \ and\ \bibinfo
  {author} {\bibfnamefont {M.-C.}\ \bibnamefont {{B}a\ nuls}},\ }\href
  {\doibase 10.1103/PhysRevLett.107.165301} {\bibfield  {journal} {\bibinfo
  {journal} {Phys. Rev. Lett.}\ }\textbf {\bibinfo {volume} {107}},\ \bibinfo
  {pages} {165301} (\bibinfo {year} {2011})}\BibitemShut {NoStop}%
\bibitem [{\citenamefont {{T}rotzky}\ \emph {et~al.}(2008)\citenamefont
  {{T}rotzky}, \citenamefont {{C}heinet}, \citenamefont {{F}\"olling},
  \citenamefont {{F}eld}, \citenamefont {{S}chnorrberger}, \citenamefont
  {{R}ey}, \citenamefont {{P}olkovnikov}, \citenamefont {{D}emler},
  \citenamefont {{L}ukin},\ and\ \citenamefont
  {{B}loch}}]{Trotzky-SuperexchangeColdAtoms-Science-2008}%
  \BibitemOpen
  \bibfield  {author} {\bibinfo {author} {\bibfnamefont {S.}~\bibnamefont
  {{T}rotzky}}, \bibinfo {author} {\bibfnamefont {P.}~\bibnamefont
  {{C}heinet}}, \bibinfo {author} {\bibfnamefont {S.}~\bibnamefont
  {{F}\"olling}}, \bibinfo {author} {\bibfnamefont {M.}~\bibnamefont {{F}eld}},
  \bibinfo {author} {\bibfnamefont {U.}~\bibnamefont {{S}chnorrberger}},
  \bibinfo {author} {\bibfnamefont {A.~M.}\ \bibnamefont {{R}ey}}, \bibinfo
  {author} {\bibfnamefont {A.}~\bibnamefont {{P}olkovnikov}}, \bibinfo {author}
  {\bibfnamefont {E.~A.}\ \bibnamefont {{D}emler}}, \bibinfo {author}
  {\bibfnamefont {M.~D.}\ \bibnamefont {{L}ukin}}, \ and\ \bibinfo {author}
  {\bibfnamefont {I.}~\bibnamefont {{B}loch}},\ }\href@noop {} {\bibfield
  {journal} {\bibinfo  {journal} {Science}\ }\textbf {\bibinfo {volume}
  {319}},\ \bibinfo {pages} {295} (\bibinfo {year} {2008})}\BibitemShut
  {NoStop}%
\bibitem [{\citenamefont {{C}hen}\ \emph {et~al.}(2011)\citenamefont {{C}hen},
  \citenamefont {{N}ascimb\`ene}, \citenamefont {{A}idelsburger}, \citenamefont
  {{A}tala}, \citenamefont {{T}rotzky},\ and\ \citenamefont
  {{B}loch}}]{PhysRevLett.107.210405}%
  \BibitemOpen
  \bibfield  {author} {\bibinfo {author} {\bibfnamefont {Y.-A.}\ \bibnamefont
  {{C}hen}}, \bibinfo {author} {\bibfnamefont {S.}~\bibnamefont
  {{N}ascimb\`ene}}, \bibinfo {author} {\bibfnamefont {M.}~\bibnamefont
  {{A}idelsburger}}, \bibinfo {author} {\bibfnamefont {M.}~\bibnamefont
  {{A}tala}}, \bibinfo {author} {\bibfnamefont {S.}~\bibnamefont {{T}rotzky}},
  \ and\ \bibinfo {author} {\bibfnamefont {I.}~\bibnamefont {{B}loch}},\ }\href
  {\doibase 10.1103/PhysRevLett.107.210405} {\bibfield  {journal} {\bibinfo
  {journal} {Phys. Rev. Lett.}\ }\textbf {\bibinfo {volume} {107}},\ \bibinfo
  {pages} {210405} (\bibinfo {year} {2011})}\BibitemShut {NoStop}%
\bibitem [{\citenamefont {{M}anoharan}\ \emph {et~al.}(2000)\citenamefont
  {{M}anoharan}, \citenamefont {{L}utz},\ and\ \citenamefont
  {{E}igler}}]{manoharan-TwoKondoMirage-Nature-2000}%
  \BibitemOpen
  \bibfield  {author} {\bibinfo {author} {\bibfnamefont {H.~C.}\ \bibnamefont
  {{M}anoharan}}, \bibinfo {author} {\bibfnamefont {C.~P.}\ \bibnamefont
  {{L}utz}}, \ and\ \bibinfo {author} {\bibfnamefont {D.~M.}\ \bibnamefont
  {{E}igler}},\ }\href@noop {} {\bibfield  {journal} {\bibinfo  {journal}
  {Nature (London)}\ }\textbf {\bibinfo {volume} {403}},\ \bibinfo {pages}
  {512} (\bibinfo {year} {2000})}\BibitemShut {NoStop}%
\bibitem [{\citenamefont {{G}omes}\ \emph {et~al.}(2012)\citenamefont
  {{G}omes}, \citenamefont {{M}ar}, \citenamefont {{K}o {W}.}, \citenamefont
  {{G}uinea},\ and\ \citenamefont
  {{M}anoharan}}]{Manoharan-ArtificialMolecularGraphene-Nature-2012}%
  \BibitemOpen
  \bibfield  {author} {\bibinfo {author} {\bibfnamefont {K.~K.}\ \bibnamefont
  {{G}omes}}, \bibinfo {author} {\bibfnamefont {W.}~\bibnamefont {{M}ar}},
  \bibinfo {author} {\bibnamefont {{K}o {W}.}}, \bibinfo {author}
  {\bibfnamefont {F.}~\bibnamefont {{G}uinea}}, \ and\ \bibinfo {author}
  {\bibfnamefont {H.~C.}\ \bibnamefont {{M}anoharan}},\ }\href@noop {}
  {\bibfield  {journal} {\bibinfo  {journal} {Nature (London)}\ }\textbf
  {\bibinfo {volume} {483}},\ \bibinfo {pages} {306} (\bibinfo {year}
  {2012})}\BibitemShut {NoStop}%
\bibitem [{\citenamefont {{G}ao}\ \emph {et~al.}(2008)\citenamefont {{G}ao},
  \citenamefont {{W}ulfhekel},\ and\ \citenamefont
  {{K}irschner}}]{PhysRevLett.101.267205}%
  \BibitemOpen
  \bibfield  {author} {\bibinfo {author} {\bibfnamefont {C.~L.}\ \bibnamefont
  {{G}ao}}, \bibinfo {author} {\bibfnamefont {W.}~\bibnamefont {{W}ulfhekel}},
  \ and\ \bibinfo {author} {\bibfnamefont {J.}~\bibnamefont {{K}irschner}},\
  }\href {\doibase 10.1103/PhysRevLett.101.267205} {\bibfield  {journal}
  {\bibinfo  {journal} {Phys. Rev. Lett.}\ }\textbf {\bibinfo {volume} {101}},\
  \bibinfo {pages} {267205} (\bibinfo {year} {2008})}\BibitemShut {NoStop}%
\bibitem [{\citenamefont {{W}a\ifmmode~\acute{s}\else \'{s}\fi{}niowska}\ \emph
  {et~al.}(2010)\citenamefont {{W}a\ifmmode~\acute{s}\else \'{s}\fi{}niowska},
  \citenamefont {{S}chr\"oder}, \citenamefont {{F}erriani},\ and\ \citenamefont
  {{H}einze}}]{PhysRevB.82.012402}%
  \BibitemOpen
  \bibfield  {author} {\bibinfo {author} {\bibfnamefont {M.}~\bibnamefont
  {{W}a\ifmmode~\acute{s}\else \'{s}\fi{}niowska}}, \bibinfo {author}
  {\bibfnamefont {S.}~\bibnamefont {{S}chr\"oder}}, \bibinfo {author}
  {\bibfnamefont {P.}~\bibnamefont {{F}erriani}}, \ and\ \bibinfo {author}
  {\bibfnamefont {S.}~\bibnamefont {{H}einze}},\ }\href {\doibase
  10.1103/PhysRevB.82.012402} {\bibfield  {journal} {\bibinfo  {journal} {Phys.
  Rev. B}\ }\textbf {\bibinfo {volume} {82}},\ \bibinfo {pages} {012402}
  (\bibinfo {year} {2010})}\BibitemShut {NoStop}%
\bibitem [{\citenamefont {{V}illain}(1977)}]{villain-JPhysFrance-1977}%
  \BibitemOpen
  \bibfield  {author} {\bibinfo {author} {\bibfnamefont {J.}~\bibnamefont
  {{V}illain}},\ }\href@noop {} {\bibfield  {journal} {\bibinfo  {journal} {J.
  Phys France}\ }\textbf {\bibinfo {volume} {38}},\ \bibinfo {pages} {385}
  (\bibinfo {year} {1977})}\BibitemShut {NoStop}%
\bibitem [{\citenamefont {{H}enley}(1989)}]{PhysRevLett.62.2056}%
  \BibitemOpen
  \bibfield  {author} {\bibinfo {author} {\bibfnamefont {C.~L.}\ \bibnamefont
  {{H}enley}},\ }\href {\doibase 10.1103/PhysRevLett.62.2056} {\bibfield
  {journal} {\bibinfo  {journal} {Phys. Rev. Lett.}\ }\textbf {\bibinfo
  {volume} {62}},\ \bibinfo {pages} {2056} (\bibinfo {year}
  {1989})}\BibitemShut {NoStop}%
\bibitem [{\citenamefont {{C}halker}\ \emph {et~al.}(1992)\citenamefont
  {{C}halker}, \citenamefont {{H}oldsworth},\ and\ \citenamefont
  {{S}hender}}]{PhysRevLett.68.855}%
  \BibitemOpen
  \bibfield  {author} {\bibinfo {author} {\bibfnamefont {J.~T.}\ \bibnamefont
  {{C}halker}}, \bibinfo {author} {\bibfnamefont {P.~C.~W.}\ \bibnamefont
  {{H}oldsworth}}, \ and\ \bibinfo {author} {\bibfnamefont {E.~F.}\
  \bibnamefont {{S}hender}},\ }\href {\doibase 10.1103/PhysRevLett.68.855}
  {\bibfield  {journal} {\bibinfo  {journal} {Phys. Rev. Lett.}\ }\textbf
  {\bibinfo {volume} {68}},\ \bibinfo {pages} {855} (\bibinfo {year}
  {1992})}\BibitemShut {NoStop}%
\bibitem [{\citenamefont {{F}riedan}(1980)}]{PhysRevLett.45.1057}%
  \BibitemOpen
  \bibfield  {author} {\bibinfo {author} {\bibfnamefont {D.}~\bibnamefont
  {{F}riedan}},\ }\href {\doibase 10.1103/PhysRevLett.45.1057} {\bibfield
  {journal} {\bibinfo  {journal} {Phys. Rev. Lett.}\ }\textbf {\bibinfo
  {volume} {45}},\ \bibinfo {pages} {1057} (\bibinfo {year}
  {1980})}\BibitemShut {NoStop}%
\bibitem [{\citenamefont {{S}achdev}(1999)}]{sachdev_qpt_book}%
  \BibitemOpen
  \bibfield  {author} {\bibinfo {author} {\bibfnamefont {S.}~\bibnamefont
  {{S}achdev}},\ }\href@noop {} {\emph {\bibinfo {title} {{Q}uantum {P}hase
  {T}ransitions}}}\ (\bibinfo  {publisher} {Cambridge University Press},\
  \bibinfo {address} {Cambridge, U.K.},\ \bibinfo {year} {1999})\BibitemShut
  {NoStop}%
\bibitem [{\citenamefont {{H}aldane}(1983)}]{PhysRevLett.50.1153}%
  \BibitemOpen
  \bibfield  {author} {\bibinfo {author} {\bibfnamefont {F.~D.~M.}\
  \bibnamefont {{H}aldane}},\ }\href {\doibase 10.1103/PhysRevLett.50.1153}
  {\bibfield  {journal} {\bibinfo  {journal} {Phys. Rev. Lett.}\ }\textbf
  {\bibinfo {volume} {50}},\ \bibinfo {pages} {1153} (\bibinfo {year}
  {1983})}\BibitemShut {NoStop}%
\bibitem [{\citenamefont {{D}ombre}\ and\ \citenamefont
  {{R}ead}(1989)}]{PhysRevB.39.6797}%
  \BibitemOpen
  \bibfield  {author} {\bibinfo {author} {\bibfnamefont {T.}~\bibnamefont
  {{D}ombre}}\ and\ \bibinfo {author} {\bibfnamefont {N.}~\bibnamefont
  {{R}ead}},\ }\href {\doibase 10.1103/PhysRevB.39.6797} {\bibfield  {journal}
  {\bibinfo  {journal} {Phys. Rev. B}\ }\textbf {\bibinfo {volume} {39}},\
  \bibinfo {pages} {6797} (\bibinfo {year} {1989})}\BibitemShut {NoStop}%
\bibitem [{\citenamefont {{C}hakravarty}\ \emph {et~al.}(1989)\citenamefont
  {{C}hakravarty}, \citenamefont {{H}alperin},\ and\ \citenamefont
  {{N}elson}}]{PhysRevB.39.2344}%
  \BibitemOpen
  \bibfield  {author} {\bibinfo {author} {\bibfnamefont {S.}~\bibnamefont
  {{C}hakravarty}}, \bibinfo {author} {\bibfnamefont {B.~I.}\ \bibnamefont
  {{H}alperin}}, \ and\ \bibinfo {author} {\bibfnamefont {D.~R.}\ \bibnamefont
  {{N}elson}},\ }\href {\doibase 10.1103/PhysRevB.39.2344} {\bibfield
  {journal} {\bibinfo  {journal} {Phys. Rev. B}\ }\textbf {\bibinfo {volume}
  {39}},\ \bibinfo {pages} {2344} (\bibinfo {year} {1989})}\BibitemShut
  {NoStop}%
\bibitem [{\citenamefont {{A}zaria}\ \emph {et~al.}(1990)\citenamefont
  {{A}zaria}, \citenamefont {{D}elamotte},\ and\ \citenamefont
  {{J}olicoeur}}]{PhysRevLett.64.3175}%
  \BibitemOpen
  \bibfield  {author} {\bibinfo {author} {\bibfnamefont {P.}~\bibnamefont
  {{A}zaria}}, \bibinfo {author} {\bibfnamefont {B.}~\bibnamefont
  {{D}elamotte}}, \ and\ \bibinfo {author} {\bibfnamefont {T.}~\bibnamefont
  {{J}olicoeur}},\ }\href {\doibase 10.1103/PhysRevLett.64.3175} {\bibfield
  {journal} {\bibinfo  {journal} {Phys. Rev. Lett.}\ }\textbf {\bibinfo
  {volume} {64}},\ \bibinfo {pages} {3175} (\bibinfo {year}
  {1990})}\BibitemShut {NoStop}%
\bibitem{OneLoopPolykovRG}
  We found the one-loop part also using Polyakov scaling. 
\bibitem [{\citenamefont {{H}amilton}(1982)}]{Hamilton_RicciFlow_1982}%
  \BibitemOpen
  \bibfield  {author} {\bibinfo {author} {\bibfnamefont {R.~S.}\ \bibnamefont
  {{H}amilton}},\ }\href@noop {} {\bibfield  {journal} {\bibinfo  {journal} {J.
  Differential Geom.}\ }\textbf {\bibinfo {volume} {17}},\ \bibinfo {pages}
  {255} (\bibinfo {year} {1982})}\BibitemShut {NoStop}%
\bibitem [{\citenamefont {{C}haikin}\ and\ \citenamefont
  {{L}ubensky}(1995)}]{ChaikinLubensky-Book}%
  \BibitemOpen
  \bibfield  {author} {\bibinfo {author} {\bibfnamefont {P.~M.}\ \bibnamefont
  {{C}haikin}}\ and\ \bibinfo {author} {\bibfnamefont {T.~C.}\ \bibnamefont
  {{L}ubensky}},\ }\href@noop {} {\emph {\bibinfo {title} {{P}rinciples of
  condensed matter physics}}}\ (\bibinfo  {publisher} {Cambridge University
  Press},\ \bibinfo {address} {Cambridge, U.K.},\ \bibinfo {year}
  {1995})\BibitemShut {NoStop}%
\bibitem [{\citenamefont {Fellows}\ \emph {et~al.}(2012)\citenamefont
  {Fellows}, \citenamefont {Carr}, \citenamefont {Hooley},\ and\ \citenamefont
  {Schmalian}}]{PhysRevLett.109.155703}%
  \BibitemOpen
  \bibfield  {author} {\bibinfo {author} {\bibfnamefont {J.~M.}\ \bibnamefont
  {Fellows}}, \bibinfo {author} {\bibfnamefont {S.~T.}\ \bibnamefont {Carr}},
  \bibinfo {author} {\bibfnamefont {C.~A.}\ \bibnamefont {Hooley}}, \ and\
  \bibinfo {author} {\bibfnamefont {J.}~\bibnamefont {Schmalian}},\ }\href
  {\doibase 10.1103/PhysRevLett.109.155703} {\bibfield  {journal} {\bibinfo
  {journal} {Phys. Rev. Lett.}\ }\textbf {\bibinfo {volume} {109}},\ \bibinfo
  {pages} {155703} (\bibinfo {year} {2012})}\BibitemShut {NoStop}%
\bibitem [{\citenamefont {Price}\ and\ \citenamefont
  {Perkins}(2012)}]{PhysRevLett.109.187201}%
  \BibitemOpen
  \bibfield  {author} {\bibinfo {author} {\bibfnamefont {C.~C.}\ \bibnamefont
  {Price}}\ and\ \bibinfo {author} {\bibfnamefont {N.~B.}\ \bibnamefont
  {Perkins}},\ }\href {\doibase 10.1103/PhysRevLett.109.187201} {\bibfield
  {journal} {\bibinfo  {journal} {Phys. Rev. Lett.}\ }\textbf {\bibinfo
  {volume} {109}},\ \bibinfo {pages} {187201} (\bibinfo {year}
  {2012})}\BibitemShut {NoStop}%
\bibitem [{\citenamefont {Chern}\ \emph {et~al.}(2012)\citenamefont {Chern},
  \citenamefont {Fernandes}, \citenamefont {Nandkishore},\ and\ \citenamefont
  {Chubukov}}]{PhysRevB.86.115443}%
  \BibitemOpen
  \bibfield  {author} {\bibinfo {author} {\bibfnamefont {G.-W.}\ \bibnamefont
  {Chern}}, \bibinfo {author} {\bibfnamefont {R.~M.}\ \bibnamefont
  {Fernandes}}, \bibinfo {author} {\bibfnamefont {R.}~\bibnamefont
  {Nandkishore}}, \ and\ \bibinfo {author} {\bibfnamefont {A.~V.}\ \bibnamefont
  {Chubukov}},\ }\href {\doibase 10.1103/PhysRevB.86.115443} {\bibfield
  {journal} {\bibinfo  {journal} {Phys. Rev. B}\ }\textbf {\bibinfo {volume}
  {86}},\ \bibinfo {pages} {115443} (\bibinfo {year} {2012})}\BibitemShut
  {NoStop}%
\end{thebibliography}

%\bibitem{OneLoopPolykovRG}
%  We found the one-loop part also using Polyakov scaling. 


\end{document}